# On Nonlinear Radial Adiabatic Pulsations of Polytropes

Mikhail I. Ivanov[*]

**ABSTRACT**

In [arXiv: 1312.1118] the author has found the formal series describing radial adiabatic pulsations of polytropes. These series include a time-dependent function, unknown yet. If a polytrope is uniform this function is obtainable; it is governed by the well-known equation of Bhatnagar and Kothari. In the present work it is shown that the pulsations of a polytrope with a near-equilibrium stratification are described by the same equation. These pulsations are nonlinear and their periods do not equal to the periods of the standard model. The periods of the vanishingly small pulsations are obtained; using them, the period-mass-radius relation is deduced. This relation is used to estimate masses of 11 chosen Cepheids with firmly known radii; all the masses found are between 5 and 11 solar masses that correlate well with the results of the stellar evolution theory.

## INTRODUCTION

It is well known that classical Cepheids are of primary importance in determination of intergalactic distances. But they are of interest not only in this. Nowadays it is determined that a Cepheid is not an anomaly but a regular stage of evolution of stars of certain masses [1]. Hence, a study of Cepheids aids progressing of the stellar evolution theory.

The stellar pulsation modelling has a long history. In 1918 Eddington linearized the equations of the adiabatic model and obtained the second-order ODE describing small harmonic oscillations near the adiabatic equilibrium of a star [2]. This model was called the standard one. But real pulsations of the most of Cepheids are distinctly nonlinear [1].

A considerable amount of time has passed before the next step in the stellar pulsation modelling was done. Zhevakin supposed real Cepheid pulsations to be self-excited ones and to be related to existence of a helium ionization zone in an interior of the star varying opacity with the temperature [3-5]. Initially he considered a star with mass to be concentrated completely in its surface (so-called one-zone model) [3]. In this way Zhevakin could reduce the Cepheid pulsation problem to ODE integration and could understand its basic properties. Later he considered linear non-adiabatic oscillations of stars involving multiple "discrete" concentric spherical gas layers [6]. It was supposed that each layer is infinitely thin but encloses a certain part of the complete mass of the star. An analogous "discrete" multilayer model was used by Aleshin for investigation of nonlinear adiabatic stellar pulsations [7]. In 1964 Aleshin applied this technique for integration of the problem of nonlinear non-adiabatic stellar pulsations [8]. His "discrete" model included 10 concentric layers in a stellar envelope and he chase an initial motion to be a small-amplitude solution of the linearized system. The pulsations found were near-sinusoidal in deeper layers but greatly diverged from the sinusoidal form in upper layers after few pulsation periods.

Emergence of high-speed computers has allowed making more detailed calculations. In 1962-1965 Baker and Kippenhahn investigated the problem of linear non-adiabatic stellar pulsations [9, 10]. Christy studied nonlinear non-adiabatic pulsations [11]. He used an initial motion of the problem in a form of high-degree polynomials in the radial coordinate. He could establish the problem spectrum to include periodic nonlinear modes of large amplitudes that were absolutely similar to limit cycles of nonlinear ODE's. It turned out that one could obtain different modes (the fundamental tune, the first overtone and so on) in accordance with a choice of an initial motion.

---

[*] A. Ishlinsky Institute for Problems in Mechanics, Moscow, Russia; *e-mail*: m-i-ivanov@mail.ru

In more recent works initial motions in forms of solutions of the linearized system were in common use (see, for example, [12, 13]). This technique seems to be natural but it is never justified. In fact, one holds that a star is to pass through a state of small linear disturbances when it becomes unstable. But this assumption seems to be too risky taking into consideration that we, in general, do not know a spectrum structure of the nonlinear problem. Moreover, it turned out that the spectra of the pulsations investigated incorporate both "softly"-excited cycles and "hardly"-excited cycles. In order to calculate the latter ones one has to use an initial motion, which is the linearized system solution of large amplitude [13].

In order to obtain further insight into this structure a rich variety of so-called one-zone models was constructed. An idea of these models is to substitute a star for a spherical layer in a stellar interior. This allows doing away with derivatives with respect to the radial coordinate. The good overviews of one-zone models are available in [14, 15]. That is why we briefly discuss only few most interesting one-zone models.

Aleshin considered an adiabatic model where a firm core was embedded by a moving uniform envelope [7]. The radial velocities obtained were asymmetric in time as in real Cepheid pulsations. Rudd and Rosenberg generalized the model of Aleshin attempting to incorporate non-adiabaticity of pulsations [16]. It was found that the model constructed has limit cycles; the computed limit cycle of $\delta$ Cephei turns out to be very close to the observed one. This model was generalized, too. Stellingwerf associated the non-adiabacity parameter of Rudd and Rosenberg having abstract nature in [16] with a stellar luminosity; as a result, an order of the system climbed to 3 [17]. Stellingwerf could obtain the light curves with hints on a secondary bump, which is observed in Cepheids having periods longer than 6 days [18]. Besides, several works also generalized the above one-zone models in the case of a slow stellar rotation [19, 20].

However, the fact of elimination of a stellar stratification sets to exercise caution in estimation of successes of one-zone models. Perhaps, as pointed out in [14], no one-zone model can describe a real dynamics of a pulsating star.

In [21] the author has found the formal series describing radial adiabatic pulsations of polytropes. These series include a time-dependent function, unknown yet. If this function were known then the problem of adiabatic stellar pulsations would be immediately solved. In this work the author finds the sought-for function for the near-equilibrium stratification case.

In the sections 1-2 we present the results derived in [21]. The original results of the present work are presented in the following sections.

## 1. FORMULATION OF THE PROBLEM

Write the basic equations of the problem in the Lagrangian variables. Let a star has a radius $a$ and a density $\rho_0$ at an initial time $t = 0$. Denoting a radius, a density, and a pressure at an arbitrary time $t$ by $r$, $\rho$, and $p$, respectively, we write the system of equations:

$$\left(\frac{\partial^2 r}{\partial t^2} + \frac{Gm}{r^2}\right)\frac{\partial r}{\partial a} + \frac{1}{\rho}\frac{\partial p}{\partial a} = 0 \qquad (1.1)$$

$$\frac{\partial m}{\partial a} = 4\pi r^2 \rho \frac{\partial r}{\partial a} \qquad (1.2)$$

$$r^2 \rho \frac{\partial r}{\partial a} = a^2 \rho_0 \qquad (1.3)$$

$$p = K\rho^\gamma \tag{1.4}$$

$$p = \frac{R_g}{\mu}\rho T \tag{1.5}$$

Here, (1.1) is the equation of motion, (1.2) is the equation for the gravity force, (1.3) is the continuity equation, (1.4) is the polytropic equation of state, and (1.5) is the equation of ideal gas; $m(a)$ is a stellar mass bounded by the sphere having the radius $a$ at the time $t = 0$, $T$ is a temperature, $G$ is the gravitational constant, $\gamma > 4/3$ is a ratio of the specific heats of the gas, $R_g$ is the universal gas constant, $\mu$ is an average molecular mass of the gas, $K$ is some constant. The equation (1.5) just allows calculating a temperature and it is not in use in the subsequent discussion.

Note that from the equations (1.2)-(1.3) it follows that the mass function is time-constant in the Lagrangian variables and it correlates with the Lagrangian coordinate $a$ by one-to-one correspondence. Physically, it is equivalent of the fact that a sequence of stellar layers does not change at all radial motions. Thus we can use $m$ as a new variable in the system (1.1)-(1.4). Then, the system (1.1)-(1.4) reduces:

$$\left(\frac{\partial^2 r}{\partial t^2} + \frac{Gm}{r^2}\right)\frac{\partial r}{\partial m} + \gamma K\rho^{\gamma-2}\frac{\partial \rho}{\partial m} = 0 \tag{1.6}$$

$$4\pi r^2 \rho \frac{\partial r}{\partial m} = 1 \tag{1.7}$$

Enter a typical mass $M$ (the complete mass of the star), a typical radius $R$ (the radius of the star) and a typical time $\Omega = \sqrt{R^3/(\gamma GM)}$, and turn to the dimensionless variables. Expressing the density from the equation (1.7) and substituting it in (1.6), we obtain the only equation for the dimensionless function $r(m,t)$:

$$krr_{mm} + 2kr_m^2 - \gamma^{-1}mr^{2\gamma-3}r_m^{\gamma+1} - r^{2\gamma-1}r_m^{\gamma+1}r_{tt} = 0 \tag{1.8}$$

Here, $k = (4\pi)^{1-\gamma} KG^{-1}R^{4-3\gamma}M^{\gamma-2}$ is the only dimensionless number of the problem.

The equation (1.8) has to be complemented by two boundary conditions and one initial condition. First, the polytrope's central density has to be finite:

$$\lim_{m\to 0}\rho(m,t) < \infty$$

This can be expressed in the dimensionless variables as:

$$\lim_{m\to 0} r(m,t)^{-2} r_m(m,t)^{-1} < \infty \tag{1.9}$$

Second, the polytrope's surface pressure has to be equal to a pressure of an interstellar medium that we suppose to be zero:

$$\lim_{m\to M} p(m,t) = 0$$

This can be expressed in the dimensionless variables as:

$$\lim_{m\to 1} r(m,t)^{-2} r_m(m,t)^{-1} = 0 \tag{1.10}$$

The initial condition is:

$$r(1,0) = 1 \tag{1.11}$$

that is, the star of the mass $M$ has the radius $R$ at the initial time $t = 0$.

## 2. TOTAL ENERGY OF THE STAR

Some important information on the problem can result from inspection of the total energy of the star. It includes the gravitational (potential), the kinetic and the internal terms $E = E_{grav} + E_{kin} + E_{int}$. In the physical variables these terms are expressed as:

$$E_{grav} = -G \int_0^M \frac{m \, dm}{r} \tag{2.1}$$

$$E_{kin} = \frac{1}{2} \int_0^M v^2 dm \tag{2.2}$$

$$E_{int} = \frac{1}{\gamma-1} \int_0^M \frac{p}{\rho} dm \tag{2.3}$$

Turn in (2.1)-(2.3) to the dimensionless variables:

$$E_{grav} = -E_0 \int_0^1 \frac{m \, dm}{r} \tag{2.4}$$

$$E_{kin} = E_0 \cdot \frac{\gamma}{2} \int_0^1 r_t^2 \, dm \tag{2.5}$$

$$E_{int} = E_0 \cdot \frac{k}{\gamma-1} \int_0^1 \left( r^2 r_m \right)^{1-\gamma} dm \tag{2.6}$$

where $E_0 = GM^2 / R$ is a typical energy.

Rearrange the term (2.6). From (1.8) we have:

$$\left( r^2 r_m \right)^{1-\gamma} = \frac{1}{2k\gamma} \frac{m}{r} + \frac{1}{2k} r r_{tt} - \frac{1}{2} r^{3-2\gamma} r_m^{-1-\gamma} r_{mm} \tag{2.7}$$

On the other hand, the following identity is valid:

$$r^{3-2\gamma} r_m^{-1-\gamma} r_{mm} = \frac{3-2\gamma}{\gamma} \left( r^2 r_m \right)^{1-\gamma} - \frac{1}{\gamma} \frac{\partial}{\partial m} \left( r^{3-2\gamma} r_m^{-\gamma} \right) \tag{2.8}$$

Substituting (2.8) in (2.7), we obtain:

$$\left( r^2 r_m \right)^{1-\gamma} = \frac{1}{3k} \frac{m}{r} + \frac{\gamma}{3k} r r_{tt} + \frac{1}{3} \frac{\partial}{\partial m} \left( r^{3-2\gamma} r_m^{-\gamma} \right) \tag{2.9}$$

From (2.4)-(2.6), in view of (2.9), we obtain the formula for the dimensionless total energy of the star in the form:

$$\frac{E}{E_0} = \frac{1}{3(\gamma-1)} \left( \int_0^1 \left[ (4-3\gamma) \frac{m}{r} + \gamma r r_{tt} + \frac{3}{2}\gamma(\gamma-1) r_t^2 \right] dm + k r^{3-2\gamma} r_m^{-\gamma} \Big|_0^1 \right) \tag{2.10}$$

This formula ignores the boundary conditions (1.9)-(1.10). To take them into consideration one has to construct analytical series for the initial boundary value problem (1.8)-(1.11) in neighbourhoods of the singular points $m=0$ and $m=1$. The calculations performed show that these series are:

$$r(m,t) = A m^{1/3} + \frac{3^{1-\gamma}}{10k\gamma} A^{3(\gamma-1)} \left( \gamma A^2 A'' + 1 \right) m + O(m^{5/3}), \quad m \to 0, \; A = A(t) \tag{2.11}$$

$$r(m,t) = B - \frac{\gamma}{\gamma-1} k^{\frac{1}{\gamma}} B^{2\left(\frac{2}{\gamma}-1\right)} \left( \gamma B^2 B'' + 1 \right)^{-\frac{1}{\gamma}} (1-m)^{1-\frac{1}{\gamma}} +$$
$$+ O\left( (1-m)^{2\left(1-\frac{1}{\gamma}\right)} \right), \quad m \to 1, \; B = B(t) \tag{2.12}$$

where $A(t)$ and $B(t)$ are some unknown time functions. From the initial condition (1.11) it follows that $B(0) = 1$.

From (2.11)-(2.12) it is easy to obtain that the term $kr^{3-2\gamma} r_m^{-\gamma}\big|_0^1$ is identically zero. Hence, the formula (2.10) for the adiabatic case is:

$$\frac{E}{E_0} = \frac{1}{3(\gamma-1)} \int_0^1 \left[ (4-3\gamma)\frac{m}{r} + \gamma r r_{tt} + \frac{3}{2}\gamma(\gamma-1) r_t^2 \right] dm \qquad (2.13)$$

## 3. EQUILIBRIUM (THE STATIONARY STATE)

The stationary solution $r_0(m)$ describes an internal structure of a fully-convective star [22]; the equation (1.8) becomes ODE, the corresponding boundary value problem has the form:

$$kr_0 r_0'' + 2k r_0'^2 - \gamma^{-1} m r_0^{2\gamma-3} r_0'^{\gamma+1} = 0 \qquad (3.1)$$

$$r_0(m) = A_0 m^{1/3} + O(m), \quad m \to 0 \qquad (3.2)$$

$$r_0(1) = 1 \qquad (3.3)$$

$$\lim_{m \to 1} r_0'(m) = \infty \qquad (3.4)$$

The condition (3.2) is a direct consequence of (1.9). The condition (3.3) is a consequence of the initial condition (1.11) in the stationary case. The boundary value problem is not overdeterminate since the dimensionless number $k$ is unknown.

Integration of the problem (3.1)-(3.4) presents difficulties because the condition (3.3) is singular. In order to circumvent this situation we use the function $r_0$ as a new variable and the variable $m$ as a new function. With $r_0' = (dm/dr_0)^{-1}$ and $r_0'' = -(d^2 m/dr_0^2)(dm/dr_0)^{-3}$ we transform the equation (3.1) to the form:

$$-kr_0 \frac{d^2 m}{dr_0^2} + 2k \frac{dm}{dr_0} - \frac{r_0^{2\gamma-3}}{\gamma} m \left( \frac{dm}{dr_0} \right)^{2-\gamma} = 0 \qquad (3.5)$$

The conditions (3.2)-(3.4) transform to the form:

$$m(r_0) = A_0^{-3} r_0^3 + O(r_0^5), \quad r_0 \to 0 \qquad (3.6)$$

$$m(1) = 1 \qquad (3.7)$$

$$m'(1) = 0 \qquad (3.8)$$

One may use other boundary conditions, which comes from the expansion in the neighbourhood of the second singular point $r_0 = 1$:

$$m(0) = 0 \qquad (3.9)$$

$$m(r_0) = 1 + O\left( (1-r_0)^{\frac{\gamma}{\gamma-1}} \right), \quad r_0 \to 1 \qquad (3.10)$$

$$m'(r_0) = \left( \frac{k\gamma}{\gamma-1} \right)^{-\frac{1}{\gamma-1}} (1-r_0)^{\frac{1}{\gamma-1}} + ..., \quad r_0 \to 1 \qquad (3.11)$$

If $\gamma = 2$ then the equation (3.5) becomes linear. In that case the boundary value problem (3.5)-(3.8) (or (3.5), (3.9)-(3.11)) has an exact solution:

$$m(r_0) = \frac{\sin \pi r_0}{\pi} - r_0 \cos \pi r_0,$$
$$k = (2\pi^2)^{-1}, \quad A_0 = \sqrt[3]{3\pi^{-2}} \qquad (3.12)$$

In the general case, we use the method of the local separation of singularities in the same way as in [21]. The numerically found solutions of the problems (3.5)-(3.8) and (3.5), (3.9)-(3.11) are identical. The results obtained are shown in the table 1.

Table 1

| $\gamma$ | $A_0$ | $k$ |
|---|---|---|
| 1.334 | 0.268 | 0.1562 |
| 7/5 | 0.352 | 0.1279 |
| 3/2 | 0.452 | 0.1038 |
| 5/3 | 0.551 | 0.0786 |
| 2 | 0.672 | 0.0507 |

## 4. EQUATION OF NONLINEAR PULSATIONS

Consider the integrand of the formula (2.13):

$$\Phi = (4-3\gamma)\frac{m}{r} + \gamma r r_{tt} + \frac{3}{2}\gamma(\gamma-1) r_t^2 \tag{4.1}$$

It is energy per unit of mass of a star. With formulae (2.11) and (2.12) write asymptotics of $\Phi$ in the neighbourhoods of the stellar centre $m = 0$ and the stellar surface $m = 1$, respectively:

$$\Phi = \left\{ \gamma A A'' + \frac{3}{2}\gamma(\gamma-1) A'^2 - (3\gamma-4) A^{-1} \right\} m^{2/3} + O(m^{4/3}), \ m \to 0 \tag{4.2}$$

$$\Phi = \left\{ \gamma B B'' + \frac{3}{2}\gamma(\gamma-1) B'^2 - (3\gamma-4) B^{-1} \right\} + ..., \ m \to 1 \tag{4.3}$$

Note that the major terms of the series obtained are identical.

Suppose that $\Phi$ is equal to the major term of (4.2) for any $m$. In the other hand, the motions are adiabatic and:

$$\partial E / \partial t = 0 \tag{4.4}$$

hence:

$$\partial \{...\} / \partial t = 0 \tag{4.5}$$

Thus we obtain the equation for the function $A(t)$:

$$\gamma A^2 A'' + \frac{3}{2}\gamma(\gamma-1) A A'^2 - DA = 3\gamma - 4 \tag{4.6}$$

where $D$ is some constant related to the total stellar energy. Note that the dimensionless number $k$ does not enter into the equation (4.6).

The equation (4.6) allows a separation of variables:

$$A'^2 = C A^{3(1-\gamma)} + \frac{2D}{3\gamma(\gamma-1)} + \frac{2}{\gamma} A^{-1} \tag{4.7}$$

where $C$ is some constant.

If $D \geq 0$ all the solutions of the equation (4.6) describe a limitless expansion or a collapse to a point, if $\gamma = 4/3$ and $D = 0$ then the equation (4.7) takes the form:

$$A A'^2 = C + \frac{3}{2} = \text{const} \tag{4.8}$$

The solution of the equation (4.8) describes a retarding limitless expansion. For this case, it is easy to extend the solution of the equation (4.6) to a whole star; the associated stellar stratification was found in [21].

For $D < 0$, the equation (4.6) has periodic solutions.

Substituting (4.7) into (4.6), we find another form of (4.6):

$$\gamma A^2 A'' + \frac{3}{2}\gamma(\gamma-1)CA^{4-3\gamma} + 1 = 0 \qquad (4.9)$$

The equation obtained is well known in a literature. For $\gamma = 5/3$, it was deduced in [23]; it describes pulsations of a uniform star. In the general case, it was derived in [24]. Aleshin considered a model where a stationary core was embedded by a uniform envelope; if a radius of the core becomes zero the model equation took the form (4.9) [7]. Rudd and Rosenberg showed that phase portraits of pulsation cycles of some solutions of the equation (4.9) in the $r\dot{r}$-plane were close to that of the classical Cepheid $\delta$ Cephei [16]. The pulsations were of this kind that an expansion phase had longer persistence than a contraction one and it was characterized by lower radial velocities. In the other hand, using of a highly simplified stratification could not lead to physically adequate results, which was the case, as was shown in [24].

Write asymptotics of $\Phi$ with (4.6)-(4.7), (4.9):

$$\Phi(m,t) = Dm^{2/3} + f(A,A') \cdot o(m^{2/3}), \quad m \to 0 \qquad (4.10)$$

$$\Phi(m,t) = D + g(A,A') \cdot o(1), \quad m \to 1 \qquad (4.11)$$

where $f$ and $g$ are some functions which forms are not essential for us now.

If $A(t) \approx \text{const}$ then $\Phi$ weakly depends on time and the condition (4.5) is approximately satisfied. One can show that the series (2.11) when $A(t) \equiv A_0 = \text{const}$ is a solution of the stationary problem (3.1)-(3.4). Analogously, the series (2.12) when $B(t) \equiv 1$ is a solution of the same problem as well. From (4.6) we find $D_0 = (4-3\gamma)/A_0$ in the first case and $D_0 = 4-3\gamma$ in the second case. Write the Cauchy problem:

$$\gamma A^2 A'' + \frac{3}{2}\gamma(\gamma-1)AA'^2 - DA = 3\gamma - 4 \qquad (4.12)$$

$$A(0) = A_0 \qquad (4.13)$$

$$A'(0) = 0 \qquad (4.14)$$

If $D = D_0$ then the solution of the problem (4.12)-(4.14) is constant. But when $D = D_0 \pm \varepsilon$, where $\varepsilon \neq 0$ is small in reference to $D_0$, the function $A(t)$ is small-amplitude periodic with some period $T$. Denominate $T_A = \lim_{\varepsilon \to 0} T(A)$. Same deduction is also true for the function $B(t)$. Analogously, denominate $T_B = \lim_{\varepsilon \to 0} T(B)$. The periods found are limit periods of the small-amplitude nonlinear adiabatic pulsations. These pulsations are not standard model motions and, as one can show, the corresponding periods are not equal. The periods for different $\gamma$ are shown in the table 2.

Table 2

| $\gamma$ | $T_A$ | $T_B$ |
| --- | --- | --- |
| 1.334 | 22.52 | 162.4 |
| 7/5 | 3.471 | 16.62 |
| 3/2 | 3.306 | 10.88 |
| 5/3 | 3.317 | 8.109 |
| 2 | 3.460 | 6.303 |

Since all $A_0 < 1$ the periods $T_B$ are always more than the periods $T_A$ for the same $\gamma$. $T_B$ is a monotonic function of $\gamma$, whereas $T_A$ is not monotonic. Minimum of $T_A$ is roughly localized near $\gamma = 3/2$. For $\gamma$, which are not close to the critical value $\gamma = 4/3$, the periods $T_A$ weakly depend on $\gamma$ and they are localized in a narrow interval of values, roughly between 3.3 and 3.5.

We suppose that there exists a continuum of nonlinear modes of the problem (1.8)-(1.11) "near" the limit solutions considered. These modes have periods, which close to $T_A$ ($T_B$). The closer to $T_A$ ($T_B$) is a period of this mode $T$, the less is a mode amplitude until it becomes zero when $T = T_A$ ($T = T_B$). Recall that contrary to linear eigenfunctions, which we can multiply by an arbitrary constant, every nonlinear mode has its own amplitude. For Cepheids, typical radius pulsation amplitudes are 10 % [1]; a period of a 0.1-amplitude oscillation is likely near the limit period $T_A$ or $T_B$ (which, in fact, is a period of the zero-amplitude oscillation). Hence, one can roughly estimate Cepheid periods by the data of the table 2.

It is likely that a stellar stratification for a 0.1-amplitude oscillation moderately deviates from the equilibrium. More fine differences between this mode and the stationary state (as a presence or an absence of nodal lines, their disposition and so on) lie outside possibilities of the method presented; note that the fine structure of the standard model oscillations is coming to light through it is preset explicitly a priori.

## 5. CLASSICAL CEPHEIDS AND THE MODEL VERIFICATION

The periods of the table 2 relate to natural hydrodynamic adiabatic oscillations. Real Cepheids are rather more complex objects and their pulsations are self-excited oscillations [3, 4, 8]. Nevertheless, it is of interest to compare the periods of the table 2 with periods of known Cepheids because the hydrodynamic adiabatic modelling usually gives a good approximation for Cepheid periods [25].

The theoretical periods depend on three parameters – a ratio of specific heats of the stellar gas $\gamma$, a radius of a star $R$ and a mass of a star $M$. Suppose that the Cepheid consists of a fully-ionized perfect gas, then, $\gamma = 5/3$.

In the model considered a stellar boundary is defined as a surface where a pressure, a stellar density and a temperature vanish. Obviously, such stars do not exist. But adiabatic models of stellar pulsations are used, and the model radii are identified with photometric radii of real stars.

Nowadays Cepheid radii are usually measured by the Baade – Wesselink method. This method has several weak points; in this connection a number of its modifications is devised. Nevertheless, results of different researchers are closely allied for several classical Cepheids. In order to make sure in this it will suffice to compare the radius estimations of [26-29].

But a situation with Cepheid masses leaves much to be desired. Different authors, even using the same mass determination method, obtain widely deviated estimations for the same stars, in addition, with huge errors. Even estimations of so-called evolutionary masses (which are calculated using only sequels of the stellar evolution theory) of different researchers deviate very widely – it will suffice to compare the evolutionary masses of [29-33]. This is due to the fact that there exist uncertainties in parameters of the stellar evolution theory itself. Besides, different mass determination methods lead to systematically disagreed results. This phenomenon is called the mass discrepancy [30, 31].

Then, we proceed as follows: write a period-mass-radius relation. We have:

$$P = T_{A(B)}\sqrt{\frac{R^3}{\gamma GM}} \approx \frac{T_{A(B)}}{54\sqrt{\gamma}}\left(\frac{R}{R_\odot}\right)^{3/2}\left(\frac{M}{M_\odot}\right)^{-1/2} \quad (5.1)$$

where $P$ is a Cepheid period in days.

From the table 3 we have $T_A = 3.317$ and $T_B = 8.109$ for $\gamma = 5/3$. From this we find two relations for two periods:

$$\left(R/R_\odot\right)^3 = 440\left(M/M_\odot\right)P_A^2 \quad (5.2)$$

$$(R/R_\odot)^3 = 74(M/M_\odot)P_B^2 \tag{5.3}$$

From the formulae (5.2) and (5.3) one can determine theoretical Cepheid masses if their periods and their radii are known. Before we note that the closely allied radii were obtained by different researchers for some classical Cepheids. These values will be considered to be firm, using them, let calculate masses of the associated stars. The results obtained are shown in the table 3. Here, only the stars, which radii are differed from one another by less than 10 % in the latest works [26-29], are used. Their radii from the work [29] are used in the calculation.

Table 3

|  | $P, d$ | $R, R_\odot$ [26] | $R, R_\odot$ [27] | $R, R_\odot$ [28] | $R, R_\odot$ [29] | $M_A, M_\odot$ |
|---|---|---|---|---|---|---|
| δ Cep | 5.370 | 42.0 | 41.9 | 42.52±1.76 | 41.6 | 5.67 |
| V Cen | 5.495 | 42.0 | 42.6 |  | 45.3 | 7.00 |
| U Sgr | 6.745 | 47.7 | 49.7 |  | 51.4 | 6.78 |
| X Sgr | 7.014 |  | 51.2 | 52.68±3.02 | 49.8 | 5.71 |
| S Nor | 9.750 | 70.7 | 65.6 |  | 66.4 | 7.00 |
| β Dor | 9.840 |  | 66.0 | 64.71±4.15 | 64.4 | 6.27 |
| ζ Gem | 10.139 |  | 67.6 | 65.24±4.37 | 64.9 | 6.04 |
| Y Oph | 17.139 |  | 100.1 | 95.6±4.5 | 93.5 | 6.32 |
| VY Car | 19.011 | 112.8 | 108.1 |  | 108.9 | 8.12 |
| RZ Vel | 20.417 | 114.8 | 114.1 |  | 111.7 | 7.60 |
| RS Pup | 41.400 | 208.0 | 193.9 |  | 197.7 | 10.25 |

One can see that all the masses $M_A$ (which are derived from the relation (5.2)) lie in the "Cepheid range" – $3-12\,M_\odot$ [25]. The masses $M_B$ (which are derived from the relation (5.3)) are vastly larger and hence they are not displayed in the table.

One can see that a rigorous dependence between mass and period is absent. It is correlated with an absence of a rigorous dependence between radius and period in the data of [29]. This fact is attributable to a large dispersion of a period-luminosity relation known for Cepheids [18].

## BRIEF DISCUSSION

An important result of the investigation pursued is that small-amplitude nonlinear pulsations of an adiabatic polytrope have periods to be different from periods of small-amplitude linear ones. Hence, it turns out that a popular practice to use small-amplitude solutions of the linearized system as initial conditions for integration of the stellar pulsation problem can leads to loss of solutions.

It is worth noting that the found pulsation periods lead to reasonable values of stellar masses for classical Cepheids. Of course, real Cepheid pulsations are not adiabatic oscillations of polytropes. But, as noted above, the adiabatic theory gives a good approximation for non-adiabatic pulsation periods [25]. Perhaps, the better results would be obtained for sinusoidal Cepheids (DCEPS) having small pulsation amplitudes but errors of a measuring of those radii were most likely to be too large.

The results obtained unambiguously demonstrate that our knowledge on a structure of a stellar pulsation spectrum is deficient.

# REFERENCES

1. C. Hoffmeister, G. Richter, and W. Wenzel, *Variable Stars*, 1985 (NY: Springer Verlag).
2. A.S. Eddington, On the Pulsation of a Gaseous Star and the Problem of Cepheid Variables, *MNRAS*, **79**, 2-22 (1918).
3. S.A. Zhevakin, On Auto-Oscillations of One Model of Cepheids, *Doklady (Reports) of the Academy of Sciences of USSR*, **58** (3), 385-388 (1947) [in Russian].
4. S.A. Zhevakin, On the Calculation of Nonadiabatic Stellar Pulsations by Use of a Discrete Model, *Soviet Astronomy*, **3**, 267-279 (1959).
5. S.A. Zhevakin, Physical Basis of the Pulsation Theory of Variable Stars, *Ann. Rev. Astron. Astrophys.*, **1**, 367-400 (1963).
6. S.A. Zhevakin, On the Pulsational Theory of Stellar Variability, part V. Multilayer Spherical Discrete Model, *Soviet Astronomy*, **3**, 389-403 (1959).
7. V.I. Aleshin, On the Asymmetry of the Radial Velocity Curve of Cepheids, *Soviet Astronomy*, **3**, 458-465 (1959).
8. V.I. Aleshin, Auto-Oscillations of Variable Stars, *Soviet Astronomy*, **8**, 154-162 (1964).
9. N. Baker and R. Kippenhahn, The Pulsations of Models of δ Cephei Stars, *Zeitschrift für Astrophysik*, **54**, 114-151 (1962).
10. N. Baker and R. Kippenhahn, The Pulsations of Models of Delta Cephei Stars, part II, *Astrophys. J.*, **142**, 868-889 (1965).
11. R.F. Christy, The Calculation of Stellar Pulsation, *Rev. Mod. Phys.*, **36**, 555-571 (1964).
12. R.J. Talbot, Jr., Nonlinear Pulsations of Unstable Massive Main-Sequence Stars, *Astrophys. J.*, **165**, 121-138 (1971).
13. D.S. King, J.P. Cox, D.D. Eilers, and W.R. Davey, Nonlinear Cepheid Pulsation Calculations and Comparison with Linear Theory, *Astrophys. J.*, **182**, 859-884 (1973).
14. J.P. Cox, *Theory of Stellar Pulsation*, 1980 (Princeton: Princeton Univ. Press).
15. A. Munteanu, *Nonlinear One-Zone Models of Stellar Pulsations*, 2003 (Ph. Thesis, Barcelona: Universitat Politècnica de Catalunya).
16. T.J. Rudd and R.M. Rosenberg, A Simple Model for Cepheid Variability, *Astron. Astrophys.*, **6**, 193-205 (1970).
17. R.F. Stellinwerf, Luminosity Variation in the One-Zone Cepheid Model, *Astron. Astrophys.*, **21**, 91-96 (1972).
18. Y.N. Efremov, Classical Cepheids, *Pulsating Stars*, ed. by B.V. Kukarkin, 1975 (New York: John Wiley & Sons).
19. E. Antonello, Linear and Nonlinear Oscillations of Simple Stellar Models, *Astrophys. Space Sci.*, **86**, 485-491 (1982).
20. D. Pricopi, Luminosity Variation in the Extended One-Zone RR Lyrae Model, *Serb. Astron. J.*, **170**, 57-64 (2005).
21. M.I. Ivanov, Two Cases of Radial Adiabatic Motions of a Polytrope with Gamma=4/3, arXiv: 1312.1118. 10 pp.
22. M. Schwarzschild, *Structure and Evolution of the Stars*, 1958 (Princeton: Princeton Univ. Press).
23. P.L. Bhatnagar and D.S. Kothari, A Note on the Pulsation Theory of Cepheid Variables, *MNRAS*, **104**, 292-296 (1944).
24. S. Rosseland, *The Pulsation Theory of Variable Stars*, 1949 (Oxford: Clarendon Press).
25. A. Maeder, *Physics, formation and evolution of rotating stars*, 2009 (Berlin-Heidelberg: Springer Verlag).
26. J. Storm, B.W. Carney, W.P. Gieren et al., The Effect of Metallicity on the Cepheid Period-Luminocity Relation from a Baade-Wesselink Analysis of Cepheids in the Galaxy and in the Small Magellanic Cloud, arXiv: astro-ph/0401211. 23 pp.
27. P. Moskalik and N.A. Gorynya, Mean Angular Diameters and Angular Diameter Amplitudes of Bright Cepheids, arXiv: astro-ph/0507576. 14 pp.

28. M.A.T. Groenewegen, The Projection Factor, Period-Radius Relation, and Surface-Brightness Colour Relation in Classical Cepheids, *Astron. Astrophys.*, **474**, 975-981 (2007).
29. H.R. Neilson and J.B. Lester, On the Enhancement of Mass Loss in Cepheids Due to Radial Pulsation, arXiv: 0803.4198. 46 pp.
30. K. Fricke, R.S. Stobie, and P.A. Strittmatter, The Masses of Cepheid Variables, *Astrophys. J.*, **171**, 593-604 (1972).
31. A.N. Cox, Cepheid Masses from Observations and Pulsation Theory, *Astrophys. J.*, **229**, 212-222 (1979).
32. W.P. Gieren, Towards a Reconciliation of Cepheid Masses, *Astron. Astrophys.*, **225**, 381-390 (1989).
33. F. Caputo, G. Bono, G. Fiorentino et al., Pulsation and Evolutionary Masses of Classical Cepheids. I. Milky Way Variables, arXiv: astro-ph/0505149. 38 pp.